\documentclass[review]{elsarticle}

\usepackage{lineno,hyperref, bm, enumerate, amsmath, amssymb}

\newcommand{\bea}{\begin{eqnarray*}}
\newcommand{\eea}{\end{eqnarray*}}
\newcommand{\bean}{\begin{eqnarray}}
\newcommand{\eean}{\end{eqnarray}}
\newcommand{\bdm}{\begin{displaymath}}
\newcommand{\edm}{\end{displaymath}}

\def\boldfacefake #1{%
        \hbox{%
                \mathsurround=0pt
                \hbox to 0.25pt{$#1$\hss}%
                \hbox to 0.25pt{$#1$\hss}%
                \hbox {$#1$}%
        }%
}
{\hspace{-0.15 cm}
\begin{Sbox}
\begin{minipage}{12 cm}
 \vspace{0.15 cm} }%
{\vspace{0.15 cm} 
\end{minipage}
\end{Sbox}
\fbox{\TheSbox}}

\newcommand{\esl}{\end{slide}}
\newcommand{\bsl}{\begin{slide}}

\newcommand{\diag}[1]{\operatorname{\rm diag}\left(#1\right)}

\newcommand{\bei}{\begin{itemize}}
\newcommand{\eni}{\end{itemize}}
\newcommand{\beq}{\begin{equation}}
\newcommand{\enq}{\end{equation}}

\newcommand{\var}[1]{\operatorname{Var} \left(#1\right)}

\newcommand {\bdot}{\hbox{\Huge .}}
\newcommand {\dotdot}{{\hbox{\Huge .}\kern-0.1667em\hbox{\Huge .}}}
\newcommand {\onedot}{1\kern-0.1667em\bdot}
\newcommand {\twodot}{2\kern-0.1667em\bdot}
\newcommand {\idot}{i\kern-0.1667em\bdot}
\newcommand {\jdot}{j\kern-0.1667em\bdot}
\newcommand {\mdot}{m\kern-0.1667em\bdot}
\newcommand {\dotj}{\kern-0.1667em\bdot\kern-0.1667em j}

\newcommand{\slopefrac}[2]{\leavevmode\kern.1em
\raise .5ex\hbox{\the\scriptfont0 #1}\kern-.1em
/\kern-.15em\lower .25ex\hbox{\the\scriptfont0 #2}}

\journal{Journal of \LaTeX\ Templates}

\begin{document}

\begin{frontmatter}

\title{New Methods for Small Area Estimation with Linkage Uncertainty}

\author[rvt]{Dario Briscolini}
\ead{dario.briscolini@uniroma1.it}
\author[focal]{Loredana Di Consiglio}
\ead{diconsig@istat.it}
\author[rvt] {Brunero Liseo\corref{mycorrespondingauthor}}
\ead{brunero.liseo@uniroma1.it}
\author[rvt]{Andrea Tancredi}
\ead{andrea.tancredi@uniroma1.it}
\author[focal]{Tiziana Tuoto}
\ead{tuoto@istat.it}
\address[rvt]{Sapienza Universit\`a di Roma, Via del Castro Laurenziano 9, Roma 00161,Italy}
\address[fox]{Eurostat, Luxembourg}
\address[focal]{Istat, Via Cesare Balbo, 16, 00184 Roma Italy }



\cortext[mycorrespondingauthor]{Corresponding author}


\begin{abstract}
In Official Statistics, interest for data integration has been increasingly growing, due to the need of extracting information from different sources. However, the effects of these procedures on the validity of the resulting statistical analyses has been disregarded for a long time. 
In recent years, it has been largely recognized that linkage is not an error-free procedure and linkage errors, as false links and/or missed 
links, can invalidate the reliability of estimates in standard statistical models. 
In this paper we consider the general problem of making inference using data that have been probabilistically linked and we explore the effect of potential linkage errors on the production of small area estimates.
We describe the existing methods and propose and compare new approaches both from a classical and from a Bayesian perspective.
We perform a simulation study to assess pros and cons of each proposed method;
our simulation scheme aims at reproducing a realistic context both for small area estimation and record linkage procedures. 
\end{abstract}


\end{frontmatter}


\section{Data integration and impact of linkage errors}
\label{diconsiglio_sec:1}

In Official Statistics, interest for data integration has been increasingly growing, due to the need of extracting information from different sources. However, the effects of these procedures on the validity of the resulting statistical analyses has been disregarded for a long time. 
In recent years, it has been largely recognized that linkage is not an error-free procedure and linkage errors, as false links and/or missed links can invalidate the reliability of estimates in standard statistical models. 
The effect of linkage errors on the calibration of linear regression models with variables observed in different sources was firstly illustrated by Neter et al. \cite{neter}. Major contributions to the development of this study can be found in Scheuren and Winkler (\cite{SW93},\cite{SW97}) and Lahiri and Larsen \cite{lahiri2005}. 
Chambers \cite{chambers09} also considers the construction of a Best Linear Unbiased Estimator and its empirical version. He also proposes a maximum likelihood estimator, providing examples with application in linear regression models, with a partial generalization to the logistic case. A possible extension to sample-to-register linkage is also suggested. 
On the Bayesian side, Tancredi and Liseo \cite{tl15} and Tancredi et al. \cite{tsl17} have proposed an integrated model with a feed-back effect in which inferential procedures for the regression are able to borrow strength from the linkage process and vice versa.

This article focuses on the effects of linking errors on the production of small area estimates. In particular we consider the case of unit-level small area methods. They apply when some auxiliary variables $X$, whose totals are known for each small area, are available for each sampled unit. Small area predictions are usually constructed using linear (or possibly generalized) mixed models expressing the survey variable $Y$ in terms of $X$.

Samart and Chambers \cite{samartchambers14} consider the effect of linkage errors on mixed effect models, extending the settings in Chambers \cite{chambers09} and suggesting estimators of the variance effects which are adjusted for linkage errors. 
In official statistics, these mixed models are largely exploited for small area estimation in order to increase the detail of information at local level. Administrative data can also be used to increase information collected in sample surveys, in order to expand auxiliary information and improve the model fitting for small area estimation. 
Linkage of external sources with basic statistical registers as well as with sample surveys can be carried out on different linkage scenarios. Di Consiglio and Tuoto \cite{diconsiglio14} performed a sensitivity analysis for different alternative linkage error scenarios in the linear and logistic regression settings. 

In this paper, we present a comparative analysis of several different estimators of the parameters of a unit-level small area model both from a classical and a Bayesian perspective. We compare the results on 
a pseudo population, where the values of the survey variable $Y$ and those of covariates $X$ are obtained from the survey on Household Income and Wealth, Bank of Italy and the person identifiers come from the fictitious population census data \cite{ESSnet2011} created for the ESSnet DI, an European project on data integration run from 2009 to 2011. The data set contains 26,625 observations and consists of 25 variables. 
 
In a classical framework, under the assumption that false matches occur only within the same small area, the linkage error affects the small area predictors via a bias on the estimation of fixed components and random effects. In addition, sample means of the covariates would also be erroneously evaluated.
Following Chambers \cite{chambers09}, we assume that sampling does not change the outcome of the linkage process and we derive an adjusted EBLUP estimator.
We also propose a Bayesian strategy where we jointly model the record linkage and the small area model using response variable and covariates available in different data sets.
We believe that the latter approach is able - in a very natural way - to
\begin{itemize}
	\item
	improve the performance of the linkage step through the use of the extra information contained in the $Y$'s (the response variable values) and the covariates $X$'s. This happens because pairs of records which do not adequately fit the small area model, say ${\cal M}$, will be automatically down-weighted in the matching process;
	\item
	allow to account for matching uncertainty in the estimation procedure related to model ${\cal M}$ involving $Y$'s and $X$'s.
	\item
	improve the accuracy of the estimators of the parameters of model ${\cal M}$ in terms of bias.
\end{itemize}

Although we present several different strategies for estimating the parameters of the small area model, we stress the fact that a fair comparison among the different methods is not possible, since they consider different sets of assumptions. In the simulation study section we will discuss these issues in detail. 

The linkage methods used in this paper refer to those implemented in \textit{RELAIS} \cite{Relais} on the frequentist side; for the Bayesian approaches we have used the methods described in \cite{tl15} and \cite{tsl17}, where categorical variables are used for the linkage procedure, while either 
continuous or categorical variable can be considered in the inferential post-linkage step, as it might be the case in small area models.

The rest of the paper is organized as follows: Section \ref{due} describe the statistical problem of linking data both from a classical and from a Bayesian perspective. 
Section \ref{tre} illustrates the different strategies of estimation in small area models. Section \ref{quattro} compares the different methods using 
a simulation setting and a realistic pseudo-population, as described above, which mimic typical data sets to be used in record linkage problems and in small area estimation as well.
We also perform a sensitivity analysis with some simulated data sets in order to assess the impact of the various assumptions in the different approaches.

\section{Linkage model and linkage errors}
\label{due}

From a statistical perspective, the operation of merging two (or more)
data sets can be important for two different and complementary reasons:
\begin{itemize}
	\item[(i)] to obtain a larger reference data set or frame, suitable to perform more accurate statistical analyses;
	\item[(ii)] to make inference on suitable statistical models via the additional information which could not be extracted from either one of the two single data sets.
\end{itemize}
If the merging step can be accomplished without errors (maybe because an error-free identification key is available and it can
be used to match units in different data sets), there are no specific consequences on the statistical procedures undertaken in both the situations. In practice, however, identification keys are rarely available and linkage between records is usually performed under uncertainty.
This issue has caused a very active line of research among the statistical and the Information Technology communities, named ``record linkage'', where
the possibility to make wrong matching decisions must be accounted for, especially when the result of the linking operation, namely the merged data set, must be used for further statistical analyses.

In order to briefly recall what record linkage is,
let us suppose we have two data sets, say $F_1$ and $F_2$, whose records respectively relate to statistical units (e.g. individuals, firms, etc.) of partially overlapping samples (or populations), say $S_1$ and $S_2$.
Records in each data set consist of several fields, or variables, either quantitative or categorical, which may be observed together with a potential amount of measurement error. 
The goal of a record linkage procedure is to detect all the pairs of units $(j,j^\prime)$, with $j\in S_1$ and $j^\prime \in S_2$, such that $j$ and $j^\prime$ actually refer to the same unit. 
If the main goal of the record linkage process is the former outlined above (case (i)), a new data set is created by merging together three different subsets of units: those which are present in both data sets, those belonging to $S_1$ only and those belonging to $S_2$ only. 
Appropriate statistical data analyses may be then performed on the enlarged data set. Since the linkage step is done with uncertainty, the efficiency of the statistical analysis may be jeopardized by $i)$ the presence of duplicate units and $ii)$ a loss of power, mainly due to erroneous matching in the merging process.

On the other hand, the latter situation (case (ii)), which is more important for the scope of this paper, is even more challenging. 
Let us denote the observed variables in $F_1$ by $(Y, W_1,W_2,\ldots,W_h)$, whereas the observed variables in $F_2$ are $(X_1, X_2, \dots, X_p, W_1,W_2,\ldots,W_h)$.
Also suppose that one is interested in performing a small area analysis in order to produce estimates of the variable $Y$ at area level, using as covariates, variables $X$'s, restricted to those pairs of records which are declared matches after a record linkage analysis based on variables $(W_1, \dots, W_h)$. 
The intrinsic difficulties in such a problem are well documented, for the linear regression case in Neter \cite{neter} and deeply discussed in Scheuren and Winkler (\cite{SW93}, \cite{SW97}), Lahiri and Larsen \cite{lahiri2005} and Chambers \cite{chambers09}.
In the regression example, it might be easily seen that the presence of false matches (that is, matching record pairs which do not actually refer to the same statistical unit) reduces the observed level of association between $Y$ and $X$ and, as a consequence, they introduce a bias effect towards zero when estimating the slope of the regression line.
Similar biases may appear in every statistical procedure and, in most of the cases, the bias takes a specific direction. As another example, when linkage procedures are used for estimating the size $N$ of a population through a capture-recapture approach, the presence of false matches may severely reduce the final estimate of $N$.

\subsection{Record Linkage: Fellegi and Sunter's approach}
\label{2.1}
The most widespread and successful theory for record linkage was proposed by Fellegi and Sunter \cite{FS1969}. We start from two lists (i.e. a register and a sample), say $F_1$ and $F_2$, of size $N_1$ and $N_2$, and we let $\Omega= F_1 \times F_2$ be the set of all possible pairs of units belonging to different data sets.

 The goal of a linkage process can be viewed as a classification problem where the pairs in 
$\Omega= ( (i,j), i \in F_1, j \in F_2)$ have to be classified into two disjoint subsets $M$ and $U$, such that $M= \{(i,j) \in \Omega: i \equiv j \}$ is the link set and $U=\Omega \setminus M$ is the non-link set.
 At the end of the linkage procedure, two possible kinds of error may occur: i) a false match or \textit{false positive}, that is a pair is declared as a link but the two records are actually referred to different units; ii) the missing match or \textit{false negative}, that is the pair is declared as a non-link but the two records are referred to the same units. 

In a more formal way, data sets $F_1$ and $F_2$ may be represented as two matrices, say $W_1$ and $W_2$
Here
$$W_i=(w_{i1}, w_{i2}, \ldots w_{i N_i}) \qquad i = 1, 2,$$
where each single $w_{ij}$
is a vector $w_{ij}=(w_{ij1},\ldots,w_{ijh})$, that is $w_{ij}$ contains the
observed values of a categorical random vector $w=(w_1,\ldots,w_h)$ whose support is
$$
\mathcal{W}=\{w_{s_1 s_2,\ldots,s_h}= (s_1,\ldots,s_h) \quad
s_1=1\ldots,k_1;\ldots; s_h=1,\ldots k_h\}.$$
Under this notation,we have
$$
M=\{ (j,j'): \textnormal{ record } j\in W_1
\textnormal{ and } j'\in W_2 \textnormal{ refer to the same unit}\},$$
and, of course, $U=\Omega \setminus M$ is the complementary set.
Notice that, in any application, no matter what is the overlapping of the two files of records, the cardinality of $U$ is always much larger than the cardinality of $M$.
The statistical model for a record linkage analysis is built upon the so called comparison vectors
$q_{j j'}= (q_{jj'1}, \cdots, q_{jj'h})$, where, in the simplest setting, 
$$
q_{jj'l}= \left \{ \begin{array}{cc} 1 \\ 0 \end{array}
\qquad \begin{array}{cc} w_{1jl}=w_{2j'l} \\
w_{1jl}\neq w_{2j'l} \end{array} \right ., \qquad \qquad l=1,\ldots,h.
$$
The comparison vectors $q_{j j'}$ are usually assumed to be independent and identically distributed random vectors with a distribution given by the following mixture density
\begin{equation}
\label{mixbern}
p(q_{j j'}|m,u,\zeta)=\zeta \prod_{l=1}^h m_l^{q_{jj' \,l}} (1-m_l)^{1-q_{jj'\,l}}+(1-\zeta)
\prod_{l=1}^h u_l^{q_{jj'\,l}} (1-u_l)^{1-q_{jj'\,l}}.
\end{equation}
In the above formula, $\zeta$ represents the marginal probability that a random pair of records belong to the same unit. In other words, $\zeta$ may be interpreted as the percentage of overlapping of the two data sets.
The quantities $m_l$ and $u_l$, $l=1, \dots,h$, are the parameters of the two multinomial distributions associated with the two set of comparisons $M$ and $U$, that is
$$
m_l=P(q_{jj' \,l }=1| j,j'\in M)
\quad \quad
u_l=P(q_{jj' \,l }=1| j,j'\in U)$$
The independence assumption of the comparison vectors $q_{jj}$'s is, strictly speaking, untenable from a probabilistic perspective. Consider the following example: after comparing record $A_1$ with records $B_1$ and $B_2$, and then record $A_2$ with $B_1$ only, the result of the comparison between $A_2$ and $B_2$ is often already known. 
Also, in the standard setting,
the key variables are assumed independent of each other.
Several extensions of this basic set-up have been proposed, mainly by introducing potential interactions among key variables, see for example Winkler \cite{winkler:95} and Larsen and Rubin \cite{laru:01}.

To test whether a given pair should be allocated to $M$ or $U$,
one may consider either the likelihood ratio
$$
\psi = \frac{P(q_{jj'}| (j,j')\in M)}{P(q_{jj'} | (j,j') \in U)}=
\frac{\prod_{l=1}^h m_l^{q_{jj'l}} (1-m_l)^{1-q_{jj'l}}}
{\prod_{l=1}^h u_l^{q_{jj'l}} (1-u_l)^{1-q_{jj'l}}},
$$
or - in a Bayesian setting - the posterior probability that a single pair is a match
$p( (j,j') \in M |q_{jj'})$.
In general, a pair of records with a likelihood ratio $\psi$ - or a posterior probability - above a fixed threshold, is declared a match. In practice, the choice of the threshold can be problematic, as illustrated, for example, in Belin and Rubin \cite{br:95}. In this context, optimization techniques may be helpful to rule out the multiple matches issue, that is the possibility that a single unit in data set $F_1$ is linked with more than one unit in data set $F_2$.
Sadinle \cite{Sadinle17} argues that such decision rules can lead to inconsistencies and proposes alternative Bayes estimates based on loss functions.

\subsection{A Bayesian perspective on record linkage}
Tancredi and Liseo \cite{tancredi} have proposed a different approach based on the direct modeling of the observed data matrices $W_1$ and $W_2$ of the key variables, rather than using the mutual comparisons. 
This way, one is able to take into account both the potential measurement error and the matching constraints.
Let $\tilde{w}_{ijl}$ be true unobserved value for the field $l$
of the record $j$ on data set $W_i$ and let $\tilde{W}_i$ be the corresponding unobserved data matrix.
We assume that
\begin{align*}
p(W_1,W_2 \vert \tilde{W}_1,\tilde{W}_2,\nu)& =
\prod_{ijl} p(w_{ijl}| \tilde{w}_{ijl}, \nu_l) \\
&=\prod_{ijl} \left[ \nu_l I(w_{ijl}=\tilde{w}_{ijl})
+ (1-\nu_l) \xi( w_{ijl})\right] .
\end{align*}
The above expression is a mixture of two components: the former is degenerate at the true value while the latter can be any distribution whose support is the set of all possible values of the variable $W_l$;
in absence of specific information, the use of a uniform distribution for the second component of the mixture is a reasonable assumption. This way, $\xi(w_{ijl})=1/k_l$.
Also notice that, in this context, $\nu_l$
represents the probability that the variable $W_l$ is observed without noise.
This model, known as ``hit and miss'', was introduced in the record linkage literature by Copas and Hilton \cite{copas:hilton}.

In order to build a model for true values $\tilde{w}_{ijl} s$ one needs to introduce a matching matrix $C$.
In particular, let $C$ be a $N_1\times N_2$ matrix whose unknown entries are either $0$ or $1$, where $C_{jj'}=1$ represents a match, $C_{j j'}=0$ denotes a non-match.
We assume that each data set does not contain replications of the same unit, so that $\sum_{j'}C _{jj'}\leq 1 $, and $\sum_{j }C _{jj'}\leq 1.$
We also assume that the joint distribution of $\tilde{W}_1$ and $\tilde{W}_2$ both depends on the entries of the matching matrix $C$ and on the probability vector
$\theta=(\theta_{s_1\dots s_h}, s_1=1\ldots,k_1;\ldots; s_h=1\ldots, k_h)$ which describes the distribution of the true values one can observe on each sample. More precisely, we assume that
\begin{equation}
\label{tildeV}
p(\tilde{W}_1, \tilde{W}_2 |C,\theta) =\prod_{j: C_{j j'}=0 \,, \forall j'}
p(\tilde{w}_{1j}|\theta) \prod_{j': C_{j j'}=0 \,, \forall j}
p(\tilde{w}_{2j'}|\theta) \prod_{j j': C_{j j'}=1} p(\tilde{w}_{1j},
\tilde{w}_{2j'}|\theta),
\end{equation}
where
$$p(\tilde{w}_{ij}|\theta)=\prod_{s_1\ldots s_h} \theta_{s_1,\dots, s_h}^{I(\tilde{w}_{ij}=(s_1,\ldots,s_h))}, $$
and
$$
p(\tilde{w}_{1j}, \tilde{w}_{2j'}|\theta)=\left\{
\begin{array}{l l}
0 & \textnormal{if } \tilde{w}_{1j} \neq \tilde{w}_{2j'}\\
\prod_{s_1\ldots s_h} \theta_{s_1,\dots, s_h}^{I(\tilde{w}_{ij}=(s_1,\ldots,s_h))} & \textnormal{if } \tilde{w}_{1j} = \tilde{w}_{2j'}
\end{array}\right.
$$
The above record linkage model is a simplified version of the one
proposed in Tancredi and Liseo \cite{tancredi}, where an additional layer - introducing a super-population model -
was added at the top of the hierarchy.
This simplest version, already used in Hall et al.\cite{hallWP} and Tancredi and Liseo \cite{tl15},
can be easily obtained by integrating out the additional layer of hierarchy,
under specific prior assumptions.
Following Hall et al. \cite{hallWP}, we also assume that the key variables are independent.
In symbols, setting $\theta_{l,s_l}=p(\tilde{w}_{ijl }=s_l|\theta_l)$, with $\theta_l=(\theta_{l 1}, \ldots, \theta_{l,k_l})$, we assume that
$$\theta_{s_1,\dots, s_h}= \prod_{l=1}^k \theta_{l,s_l}.$$

To complete the model we need to specify a prior distribution for the matching matrix $C$ and prior distributions for the parameters $\nu_l$ and $\theta_l$, $l=1,\ldots, h$. For these
latter quantities the standard assumptions of independent Beta distributions for the probabilities $\nu_l$ and independent Dirichlet distributions for the vectors $\theta_l$ can be adopted. Regarding $C$, the prior can be elicited in two stages. First, we elicit a prior distribution $p(t)$, $t = 0, 1, 2, \dots N_1 \wedge N_2$ on $T$: ``number of matched pairs in the two data sets''. At this stage, the researcher can easily collect information, looking at previous experiences or at the statistical characteristics of the data sets (e.g. if the
two data sets refer respectively to a census and a sample, we can expect a
large number of matched pairs). At the second stage we define a conditional prior
distribution for the configuration matrix $C$ {\em given} the number of matches.
We take the natural noninformative choice of a uniform conditional prior on the set
$
C^{(t)}= \{ C : \sum_{j j'} C_{j,j^\prime}= t \big \}.
$

The model just outlined cannot be analyzed in a closed form and simulation from
the posterior distribution is necessary.
In particular, we have implemented a Metropolis within Gibbs algorithm where
the updating of parameters $\nu_l$ and $\theta_l$ can be easily performed by simulating from their respective full conditional distributions, for $l=1, \dots, h$. On the other hand, the updating of the matching matrix $C$ and the true values $\tilde{W}_1$ and $\tilde{W}_2$ is jointly obtained.
In particular, we adopt a Metropolis-Hastings step by proposing a new matching matrix $C$, which is obtained
by adding or deleting one matches or switching two already existing matches. Conditionally
on the acceptance of the proposed value for $C$, a Gibbs step is used for the updating of the elements of $\tilde{W}_1$ and $\tilde{W}_2$.
This Metropolis step can be easily adapted to specific situations which we will discuss in the application section. For example, it might be the case that 
the data set $F_1$ is a subset of $F_2$ so that we already know that the number 
of matches is exactly $N_1$. In this case the prior over $C$ will be restricted on those matrices with exactly $N_1$ matches and the Metropolis step will only propose a permutation of the matches or a simultaneous addition and deletion of matches.
Details of the algorithm can be found in Tancredi and Liseo \cite{tl15}.

Finally, in order to produce a point estimate of the matching configuration $C$, one can use the following - rather natural - strategy:
$$ \widehat{C}_{ij}=\left\{
\begin{array}{c c}
1 & \textnormal{if } p(C_{ij}=1|W_1,W_2) \geq \frac12 \\
0 & \textnormal{otherwise} \\
\end{array}
\right. .
$$
The above estimator is not the only possibility. Sadinle \cite{Sadinle17} proposed different ``point estimators'' of the $C$ matrix based on a more general class of loss functions

\section{Small area estimation based on unit linear mixed model}
\label{tre}

%
When the sample sizes within some domains are moderate or small, standard estimators are often not reliable enough to produce estimates at a finer level of (geographical) detail; for a general review on this topic, see Rao and Molina \cite{rao}.
The empirical best linear unbiased predictor (EBLUP) based on a unit level model was firstly proposed by Battese et al. \cite{battese},
to improve the reliability of estimators by exploiting the relationship between the 
target variable and the auxiliary variables.

\subsection{The unit linear mixed model }
\label{diconsiglio_subsec:3.1}
\noindent
Suppose that the population units can be grouped in $D$ areas or domains, 
let $Y$ be the target variable and $X$ be auxiliary variables observed on the same units.
Assume a linear mixed relationship between the target variable and the covariates
\begin{equation}
\label{unit}
y_{id}=X_{id}^T\beta+u_d+e_{id}, \enspace i=1,\dots, N_d, \enspace d=1,\dots, D,
\end{equation}
where $\beta$ is a $p$-dimensional vector of fixed regression coefficients and $u_d$, $d=1,\ldots, D$, are the i.i.d. random variables related to the specific area or domain contributions, with $\mathbb{E}(u_d)=0$ and $\var{u_d}=\sigma^2_{u}$ and i.i.d. errors $e_{id}$ with $\mathbb{E}(e_{id})=0$ and $ \var{e_{id}}=\sigma^2_e$.
In matrix notation
$$ Y=X\beta+Zu+e, $$
where $Z$ is the area design matrix, $Z=\textnormal{Blockdiag}(Z_d=1_{N_d}; d=1, \cdots, D)$.
The total variance is then 
$\var{Y}=V=\sigma^2_u ZZ^T+\sigma^2_e I$ or $V=\diag{V_1, \dots, V_D}$, with
$V_d=\sigma_e^2I_{N_d}+\sigma_u^2Z_dZ_d^T$.
When $\sigma^2_u$ and $\sigma^2_e$ are known, the BLU predictor of a small area mean $\bar Y_d$, is given by
\begin{equation}
\label{eblup}
\hat {\bar Y}_d^{BLUP} = \frac1{N_d} \left( \sum_{i \in \Psi_d} y_{id} + \sum_{i \notin \Psi_d} \hat y_{id}^{BLUP} \right)
\end{equation}
where $\hat y_{id}^{BLUP} = X_{id}^T \hat \beta + \hat u_d$ with
$$
\hat \beta=(X^TV^{-1}X)^{-1}X^TV^{-1}y
$$ 
$\hat u = \sigma_u Z^T V(y-X\hat \beta)$, and $\Psi_d$ is the subset of units in domain $d$ which were actually sampled.

An EBLUP is obtained by plugging the estimates $\hat{\sigma}_u$ and $\hat{\sigma}_e$ in the previous expressions. Estimation strategies for estimating $\hat{\sigma}_u$ and $\hat{\sigma}_e$ are described in \S \ref{sec_varcomponents}.

The mean squared error (MSE) of the standard EBLUP estimator is given by
\begin{equation}
\label{MSE}
MSE (\bar{Y}_d^{EBLUP}) \approx g_{1d}(\sigma^2_e{,}\sigma^2_u)+g_{2d}(\sigma^2_e{,}\sigma^2_u)+g_{3d}(\sigma^2_e{,}\sigma^2_u)
\end{equation}
see Prasad and Rao \cite{prasadrao}. 
The $g$ terms are, respectively,
$$
g_{1d}(\sigma^2_e{,}\sigma^2_u)=(1-\phi_d)\sigma^2_u
$$
$$
g_{2d}(\sigma^2_e{,}\sigma^2_u)=(\bar{X_d}-\phi_d\bar{x_d})^T(X^T V^{-1} X)^{-1}(\bar{X_d}-\phi_d\bar{x_d})
$$
where $\phi_d= \sigma^2_u/(\sigma^2_u+\sigma^2_e/n_d)$ and
$$
 g_{3d}(\sigma^2_e{,}\sigma^2_u)= (\sigma^2_u/ n^{−2}_d +n_d\sigma^2_e)^{-3}
 \sigma_e^4Var(\hat\sigma^2_u)+ \sigma_u^4Var(\hat\sigma^2_e)
 -2\sigma_e^2 \sigma_u^2Cov(\hat\sigma^2_u{,} \hat\sigma^2_e),
$$
see Rao \cite{rao}, Chapter 7, for details about the component $g_3$ when the variance components are estimated with ML.
The Prasad and Rao's \cite{prasadrao} proposal for the estimation of the MSE is given by 
\begin{equation}
\label{mse}
mse (\bar{Y}_d^{EBLUP}) \approx g_{1d}(\hat\sigma^2_e{,}\hat\sigma^2_u)+g_{2d}(\hat\sigma^2_e{,}\hat\sigma^2_u)+2g_{3d}(\hat\sigma^2_e{,}\hat\sigma^2_u)
\end{equation}
It is possible to obtain an estimate of the MSE using alternative techniques, such as bootstrap and jackknife.



\subsection{The unit linear mixed model under RL: the classical approach.}
\label{author_subsec:2.2}
Here we consider the case where the covariates $X$ and the target variable $Y$ are not observed on the same data set: for example, they have been obtained by linking a sample with a register list; in this situation, the plain use of the previous described techniques may produce strongly biased estimates.

Following Chambers \cite{chambers09} and Samart and Chambers \cite{samartchambers14},
let $y_{id}^*$ be the value of the response variable observed on unit $i$, matched with the value $X_{id}$ .

\noindent
Let $Z_2$ be a blocking variable that partitions both registers so that linkage errors may only occur within the groups of records defined by the distinct values of this variable.
In order to simplify the notation, we assume that blocks coincide with the actual domains. This implies, here, that $Z_2$ simply represents the domain indicator. 
We assume that $Z_2$ is measured without error on both the $Y$-register and the $X$-register.
An exchangeable linkage errors model can be defined by assuming that the probability of correct linkage is the same for all records in a domain.
We take the following standard assumptions (Chambers \cite{chambers09}):
\begin{enumerate}
\item the linkage is complete, i.e. the $X$-register and Y-register refer to the same population and have no duplicates, so the smallest $Y$-register is contained in the largest $X$-register;
\item the linkage is one-to-one between the $Y$ and $X$ registers; 
\item the linkage errors model is exchangeable within domains.
\end {enumerate}
Then, for each area $d$, the observed response vector may be considered a permutation of the true one, say 
$Y^*_d=A_dY_d$, where $A_d$ is a random permutation matrix such that
$\mathbb{E}(A_d \vert X)=G_d$. 
Set 
$$P(a^d_{ii}=1|X)=P(\textnormal{correct\enspace linkage})= \lambda_d$$ { and } 
$$P(a^d_{ij}=1|X)=P(\textnormal{incorrect\enspace linkage})=\gamma_d;$$
then the expected value can be written as:
\begin{equation}
\label{lam}
G_d = (\lambda_d -\gamma_d) I_{n_{d}} + \gamma_d 1_{n_{d}} 1^T_{n_{d}}.
\end{equation}
As in Chambers \cite{chambers09}, $1^T_{n_{d}}A_d =1^T_{n_{d}}$ and $A_d1_{n_{d}} =1_{n_{d}}$ thus,  $1^T_{n_{d}}G_d =1^T_{n_{d}}$ and $G_d1_{n_{d}} =1_{n_{d}}.$ That is, \eqref{lam} implies 
$$\lambda_d  + (n_{d} - 1)  \gamma_d= 1 $$
$$ \gamma_d=\frac{1-\lambda_d }{n_d-1}, $$
so, the first order properties of the linkage mechanism are completely specified by the parameters $\lambda_d$.
The values of the $\lambda_d$'s can be estimated, as suggested in \cite{kim}, using the
correctly linked/incorrectly linked status of some randomly sub-sampled linked records in sample in each domain.

Samart and Chambers \cite{samartchambers14} proposed a ratio-type corrected estimator for $\beta$
\begin{equation}
\tilde\beta_R=(X^T V^{-1}G X)^{-1} X^T V^{-1} y^{*},
\end{equation}
where $G= \diag{G_1, \dots, G_D}.$
Then, by exploiting the relationship between $y^*$ and $X$, a BLU estimator can be derived as
\begin{equation}
\tilde\beta_{BLUE}=(X^T G^T\Sigma^{-1} G X)^{-1} X^T G^T \Sigma^{-1} y^{*},
\end{equation}
which takes into account the derived variance of the observed $y^*$
\begin{equation}
\textrm{Var}\left (Y^\ast \right )=\Sigma= \sigma^2_u K +\sigma^2_eI + \tilde{V},
\end{equation}
where
\begin{equation}
\tilde{V} = \diag{\tilde{V}_1, \tilde{V}_2, \dots \tilde{V}_D},
\end{equation}
$\tilde{V}_d = \var{A_d X_d \beta}$ which is approximated by 
$$ 
\tilde{V}_d \approx \diag{(1-\lambda_d)(\lambda_d (f_{id}-\bar f_d)+\bar f_d^{(2)}-\bar f_d^2); i=1, \dots, n_d},
$$
where, for each domain $d$, $f_{id} = X_i\hat \beta$, restricted to those units in domain $d$, $\bar{f}$ and $\bar{f}^{(2)}$ are the means of $f_i$'s
and their squares respectively. Finally, $K$ is a function of the domain sizes and the vector of $\lambda$'s (Samart and Chambers \cite{samartchambers14}).

\subsection{Estimation of variance components (ML)}
\label{sec_varcomponents}
\noindent
As $\sigma_u$ and $\sigma_e$ are unknown,
they have to be estimated; usual strategies include the method of moments, maximum likelihood (ML) or restricted ML (Harville \cite{Harville1977}).
Here we confine ourselves to ML, and we assume a multivariate normal model.
In general, there are no closed form expressions for the variance component estimators.
Samart and Chambers \cite{samartchambers14} use the method of scoring as an algorithm to obtain the estimators.
In the standard case, i.e. when the variables are recorded on the same sample, one has $ y\sim N(X\beta;V)$; in the record linkage case, recall that
$ y^*\sim N(G f;\Sigma)$. The scoring algorithm can be applied on the derivatives of the previous likelihood. An estimate of $\beta$ is then obtained by replacing the variance components estimates, and an iterative process is usually needed. 

\subsection{Small area estimation under linkage errors}
\noindent 
For the purpose of small area estimation, the usual scenario 
to be considered is the linkage of a sample with a larger register. Here we assume that the register is complete, i.e. neither duplicates or coverage issues occur.
This setting is considered in Chambers \cite{chambers09}. 
Following this framework, we also assume that the record linkage process is independent of the sampling process.
Chambers \cite{chambers09} assumes that an hypothetical linkage can be performed before the sampling process.
Under these conditions, the variance component matrices $G$, $V$ and $\Sigma$ only depend on the domain variables and linkage errors, so the use of sampling weights is not really needed.
Besides these assumptions, as specified in section \ref{author_subsec:2.2}, we assume an exchangeable linkage errors model.

This implies that 
$
\hat {\bar Y}^{*} = \hat {\bar Y}
$,
so one can exploit the distribution of $Y^*$ in order to obtain the EBLU predictor
\begin{equation}
\label{eblup*}
\hat {\bar Y}_d^{*BLUP} = \frac1{N_d} \left( \sum_{i \in \Psi_d} y^*_{id} + \sum_{i \notin \Psi_d} \hat y_{id}^{BLUP} \right)
\end{equation}
where 
$\hat y_{id}^{BLUP}= G X \tilde\beta_{BLUE}+ \hat u_d$, with 
$$\hat u = (\hat{u}_1, \dots \hat{u}_D) =\sigma_u Z^T \Sigma^{-1}(y^\ast - G X\tilde \beta_{BLUE}).$$ 
For computational ease, the sum of the predicted values of non sampled units can be obtained as the difference of the population total predicted values and the sum of the sample predicted values. The EBLU predictors are obtained by replacing the estimators of the regression coefficients and variance components in (\ref{eblup*}).
A key aspect for the evaluation of the small area estimator in real cases applications is the estimation of its MSE. For the proposed small area estimator derived on the distribution of the $y^*$, even under the assumption of known record linkage errors and consequently known $G$ (i.e. not introducing additional element of variability to the standard case), the structure of $\var{y^*}=\Sigma$ is far more complex than in the standard linear mixed model setting described above as the it depends also through $\tilde V$ on $\beta$ . Consequently the structure of the MSE of $\hat {\bar Y}^{*EBLUP}$ require additional components. Moreover in practice the linkage errors are unknown, and their estimation will require the introduction of an additional source of uncertainty. Research for a new proposal for the mse of $\hat {\bar Y}^{*EBLUP}$ is needed.

\subsection{The unit linear mixed model under RL: the Bayesian approach.}
\label{usae:ba}
From a Bayesian perspective, there are no theoretical complications in adapting the integrated model proposed by Tancredi and Liseo \cite{tl15} to a small area framework.
In the following, we will make distributional assumptions which matches those described in \S \ref{author_subsec:2.2} in order to make valid comparisons.

We then assume the usual standard unit-level model (\ref{unit}), and we also suppose that both the random effects and the stochastic terms of the models are independent Gaussian random variables; in particular
$$u_d \vert \sigma_u^2 \stackrel{\textnormal{iid}}{\sim} N(0,\sigma_u^2), \mbox{ and } e_{id}\vert \sigma_e^2 \stackrel{\textnormal{iid}}{\sim}N(0, \sigma_e^2), \quad i=1, \dots, n_d;\, d=1, \dots, D.
$$

We also assume that the mean vector of the auxiliary variables for the generic area $d$, namely $\bar{\bm X}_{d}=\sum_{j=1}^{N_{d}}{\bm{x_{d,j}}},$ is known. 
Alternatively, if the domain population sizes $N_d$ are large enough, one can state that the small area means are approximately equal to
$${\mu}_d=\bar{\bm X}_{d.}^\prime \bm{\beta}+ u_d.$$
Assume that, as in the previous section, we start from data sets $F_1$ and $F_2$, both being samples of size $N_1$ and $N_2$, respectively.
Let $N_{1d}$ be the number of units belonging to domain $d$ and observed in $F_1, d=1, \dots, D.$ 
We observe, on data set $F_1$, the quantities
$(Y_{1d},W_{1,j,1},W_{1,j,2},\dots,W_{1,j,h}), d=1,2,\dots, D$; $j=1,2,\dots, N_{1d}$, and $\sum_{d=1}^{D}N_{1d}=N_1$.
Similarly, on data set $F_2$ we observe 
$(W_{2,j,1},W_{2,j,2},\dots,W_{2,j,h},X_{2,j,1},X_{2,j,2},\dots,X_{2,j,h}),\\ j=1,2,\dots, N_{2d}$, where $N_{2d}$ is the number of units belonging to domain $d$ in list $F_2$ and $\sum_{d=1}^{D} N_{2d}=N_2.$ 

Regarding the matching matrix $C$, its parameter space is restricted by the additional and reasonable constraint that false links may only occur within the same domain.
Then we assume that $C$ is block diagonal. In other words, linkage uncertainty concerns only single domains, and we assume that two units belonging to different areas cannot be matched.
This restriction allows us to separately deal with each single domain: this results in considering $D$ different $C_d$ matrices, $d=1, \dots, D$. 
It must be stressed, however, that these assumptions are relatively weaker than those required in \S~\ref{author_subsec:2.2}. In this case there are no exchangeability restrictions, and the posterior estimate of $C$ heavily relies on the observed key variables.
The Bayesian model is then completed with the elicitation of a prior distribution.
We assume standard priors on the parameters of the small area model. In particular we assume that $\beta, \sigma_e$ and $\sigma_u$ are mutually independent. Then we take an improper uniform prior for the location parameter vector $\bm{\beta}$, and an Inverse Gamma density for the variance component $\sigma_e^2$, that is 
$$\sigma_e^2 \sim IG(a_e,b_e),$$
with small values for the hyperparameters.
The choice of the prior of $\sigma_u$ is a more critical issue. Rao and Molina (\cite{rao}) suggest the use of another Inverse Gamma density in order to keep the model conditionally conjugate and, consequently, amenable to a straightforward Gibbs sampler.
However, Gelman (\cite{gel06}) noticed that, when this prior is used in its weakly informative version, that is setting $a_u=b_u=\varepsilon$ with $\varepsilon$ very small, the final posterior may be very sensitive to the value of $\varepsilon$, especially when the ``true'' value of $\sigma_u$ is very small and the number of domains is not large. This happens because, as $\varepsilon \to 0$, the resulting joint posterior would be improper. 
Gelman's (\cite{gel06}) alternative suggestion is then the use 
of an improper uniform prior over the standard deviation $\sigma_u$. This implies an improper prior for $\sigma^2_u$, which is proportional to $\sigma_u^{-1}$ and which produces a proper posterior, provided that the number of domains is larger than 3. 

The goal of a Bayesian analysis is the production of a sample from the joint posterior distribution of the above parameters and those related to the record linkage part, that is 
$$\pi(C,{\beta},\bm u, \sigma_u^2, \sigma_e^2 \vert W_1, W_2, Y_1,X_2),$$
where $Y_1$ is the vector of responses of the survey variable recorded in $F_1$ and $X_2$ is the set of covariates in $F_2$ and $\bm u=(u_1,u_2,\dots, u_D)'$. 
To sample from this distribution we adopt a straightforward Gibbs sampling with a Metropolis step which is necessary to propose values from the full conditional distribution of the matching matrix $C$, as described in \S~2.2.
However, in this specific framework, the algorithm must be tailored in such a way that the proposed values are consistent with the information that a data set is a subset of the other; this implies that one knows in advance that the total number of links must be exactly $N_1$: consequently, the range of possible proposals for moving the chain around the parameter space of $C$ is restricted to $0$/$1$ matrices $C$ of size $N_1 \times N_2$ such that there are exactly $N_1$ entries equal to $1$, each row of the matrix has a single $1$ and no more than one entry can be equal to 1 in each column of the proposed $C$. This implies that only ``switching moves'' between columns of $C$ are allowed.

For a given value of $C$, the other full conditional distributions belong to well-known families, independently on which prior is used on $\sigma_u$, either a uniform prior or an Inverse Gamma on $\sigma^2_u$. The implementation of a Gibbs algorithm (conditional on $C$) can be found in Rao and Molina, chapter 10 \cite{rao}. 

In our record linkage framework, two alternative estimation strategies can be envisaged.
\begin{enumerate}[a.]
\item Feedback strategy: the algorithm produces a sample from the joint posterior distribution of the parameters of the record linkage and of the small area model together. This allows a feedback effect: not only the small area model depends on the selected matches, but even the selection of potential links will depend on the information carried by the small area model.
\item Non-feedback strategy: The record linkage part of the model obviously affects the small area part; however the reverse does not hold: in practice, we perform a Gibbs sampling for $(\beta, \sigma_u, \sigma_e, \bm{u})$ for each single $C$ generated by the algorithm and retain the last value of the chain.

\end{enumerate}

\section{Results on simulated data}
\label{quattro}

\noindent
In this Section we describe a paradigmatic application, where we have used the fictitious population census data \cite{ESSnet2011} created for the European Statistical System Data Integration project, (ESSnet DI), and the micro-data from the Survey on Household Income and Wealth, Bank of Italy, (SHIW), freely available in anonymous form. Specifically, the ESSnet population, which comprises over 26,000 records with name, surname gender and date of birth, has been augmented by adding two new variables representing the annual income and the area domain. The values of these two variables have been drawn from the SHIW data set; in particular, the domain comprises 18 areas resulting from the aggregation of the Italian administrative regions. Table {\ref{ESSPOP}} shows some records from this population register. 

To perform a realistic record linkage and small area estimation exercise the augmented ESSNET data set has been further modified by perturbing the potential linking variables (names, gender and date of birth) via the introduction either of missing values and typos. 
Moreover, from the perturbed data set we have removed the income variable, and we have added the corresponding value of the consumption variable resulting from the SHIW data set.
A list of records from this perturbed population is shown in Table \ref{ESSPOPMOD}, as an example.

\begin{table}
\begin{footnotesize}
\caption{A sample list of records from the population register}
\label{ESSPOP}
\begin{tabular}{l l l c r r r c r}
\hline
Identifier & Name &	Surname	 & Gender & \multicolumn{3}{c}{ Date of birth}& Domain & Income\\
 & & & & Day & Month & Year & & \\
\hline
DE03US003001	& NATHAN &	RUSSELL &	M &	11	& 11	& 1934	& Area1 & 6500\\
DE03US013003	& CHARLOTTE	& JONES &	F	&26 &	4	& 1974	& Area1 & 22000\\
EX985AF008003	& OWEN &	LLOYD	M	& M & 9 &	4	& 1976 & 	Area2 & 20000\\
EX985AF015002	& EVELYN &	THOMPSON	& F	& 12	& 12 &	1990 & Area2 & 17703\\
HR167XE022003	& MACEY & SHAW	 & F & 6 & 2 & 1982 & 	Area3 & 28264\\
HR167XE027001	& OLLIE & JONES & M & 21 & 4 & 1951 & 	 Area3 & 25766\\
LS992DB012005	& OLIVIA &	ANDERSON &	F &	28 &	10 &	1995 &	Area4 & 20800\\
M141DQ001002	& MILLIE &	JAMES & F & 24 &	11 &	1972 & Area4 & 4990\\
\hline
\end{tabular}
\end{footnotesize}
\end{table}

\begin{table}
\begin{footnotesize}
\caption{A sample of list of records from the perturbed population.}
\label{ESSPOPMOD}
\begin{tabular}{l l l c r r r c r}
\hline
Identifier & Name &	Surname	 & Gender & \multicolumn{3}{c}{ Date of birth}& Domain & Consumption\\
 & & & & Day & Month & Year & & \\
\hline
DE03US003001	& NATHAN &	RUSSELL &	M &	11	& 11	& -	& Area1 &5583\\
DE03US013003	& CHARIOTTE	& JONES &	F	&26 &	4	& 1974	& Area1 & 19266\\
EX985AF008003	& OWEN &	LLOYD 	& M & 9 &	4	& 1976 & 	Area2 & 11636\\
EX985AF015002	& EVELYN &	THOMPSON	& F	& 12	& 12 &	1990 & Area2 & 16323\\
\hline
\end{tabular}
\end{footnotesize}
\end{table}

\noindent
In practice, in order to compare the various methodologies, 100 replicated samples of size 1000 have been independently and randomly selected without replacement from the perturbed population. 
Each sample has been linked to the register population by using, as key-variables, \textit{Day and Year of Birth} (with respectively 31 and 101 categories) and \textit{Gender}; the \textit{Domain} played the role of the blocking variable. The aim of the linkage process is the calibration of a small area model using the consumption as target variable and the income as covariate. Table \ref{NUMAREA} shows the population sizes for each domain and the corresponding average sample sizes. Notice that some areas comprise a very small number of records at the sample level. 

\begin{table}
\begin{footnotesize}
\caption{Population and sample size in the domains }
\begin{center}
	\label{NUMAREA} 
\begin{tabular}{l r r}
\hline
Domain & Population Size & Average sample size \\
Area1 & 2880 & 107 \\
Area2 & 2302 & 88\\
Area3 & 2443 & 92\\ 
Area4 & 2404 & 92 \\
Area5 & 314 & 11\\
Area6 & 255 & 10\\ 
Area7 & 113 & 4 \\
Area8 & 296 & 12\\
Area9 & 488 & 18\\ 
Area10 & 490 & 18 \\
Area11 & 106 & 4\\
Area12 & 421 & 16\\ 
Area13 & 231 & 9 \\
Area14 & 2840 & 107\\
Area15 & 2915 & 110\\ 
Area16 & 2325 & 87 \\
Area17 & 2354 & 87\\
Area18 & 3448 & 130\\
\hline
 \end{tabular}
\end{center}
\end{footnotesize}
\end{table}

\noindent
The classical version of the probabilistic record linkage model (\cite{FS1969}, \cite{Jaro1989}) has been implemented by means of the batch version of the software \textit{RELAIS} \cite{Relais}. The linkage procedure resulted, on average across 100 replications, on 957 declared matches; the probability of false link was close to 0.14 and the probability of missing link was about 0.04.
In each simulation we have considered as links those pairs of records whose posterior probability of being a match was larger than $0.5$. The posterior probability has been computed as 
$$
\frac{\zeta \psi}{1 - \zeta + \zeta \psi},
$$
where $\psi$ is the likelihood ratio defined in \S~\ref{2.1} and $\zeta$ is the estimated probability that a random pair of records belong to the same unit, introduced in \S~\ref{2.1} formula (\ref{mixbern}). 
The Bayesian version of the record linkage procedure has been implemented following the lines described in \S~\ref{usae:ba}; see also \cite{lt-11}, \cite{tancredi} and \cite{tl15}. Also in this case we have considered matches those pairs with a posterior probability, computed via the MCMC algorithm, higher than $0.5$
 
The main goal of this section is to relatively compare the statistical performance of the different estimators of the regression coefficients
of the mixed linear model describing the small area set up. We have considered the following estimators:

\begin{enumerate}
\item[A.] the EBLUP with $X$ and $Y$ observed on the same data set, i.e. no linkage step is considered in this setting. It should be considered as the gold standard for any comparisons;
\item[B.] the EBLUP restricted on the subset of linked records. This implies a reduction of the sample size due to missed links; however, we do not introduce linkage errors, and no false link are considered;
\item[C.] a na\"{\i}ve EBLUP, restricted on the subset of linked records, and considering $X$ and $Y$ observed on two different data sets. No adjustment for linkage errors is considered.
\item[D.] the adjusted EBLUP estimator, as in formula (10).
\item[A$^\ast$] the Bayesian version of strategy $A$: in practice a hierarchical Bayesian small area model with vaguely informative priors on the hyperparameters, as illustrated in \cite{rao}, chapter 10.

\item[C$^\ast$] the Bayesian version of strategy $C$: again a hierarchical Bayesian small area model built upon a point estimate of the matching matrix $C$.

\item[E.] the posterior mean of the regression coefficients $\bm{\beta}$ using a Bayesian approach for the linkage step based only on the key variables $W_1, \dots, W_h$ (no feed-back effect).
\item[F.] the posterior mean of the regression coefficients $\bm{\beta}$ using a Bayesian approach with both the key variables and the regression variables $X$ and $Y$. In this case there is a feedback effect which makes the posterior distribution of the matching matrix $C$ also depending on $X$ and $Y$.
\end{enumerate}

\begin{table}
	\label{one1}
	\begin{center}
		\caption{Comparison of different estimators of the regression coefficients $(\beta_0, \beta_1)$: first row reports the ``true'' estimates based on the entire population. Each other row reports mean and standard deviation of the various estimators over 100 repeated sampling of size 1000}
		\label{author_tab:2} 
		
		\begin{tabular}{p{3cm}p{2cm}p{2cm}p{2cm}p{2cm}}
			\hline\noalign{\smallskip}
			Estimates & Intercept &Sd Intercept & Slope &Sd Slope \\
			Population & 3.576 & --- & 0.538 & --- \\
			Estimates A & 3.057 & 1.412 & 0.565 & 0.070 \\
			Estimates B & 3.030 & 1.552 & 0.567 & 0.077 \\ 
			Estimates C & 5.224 & 1.367 & 0.450 & 0.073 \\
			Estimates D & 3.008 & 1.633 & 0.567 & 0.086 \\
 Estimates $A^{\ast}$ & 3.045 & 1.399 & 0.566 & 0.070\\
 Estimates $C^{\ast}$ & 3.749 & 1.592 & 0.533 & 0.081\\
			Estimates E & 4.099 & 1.285 & 0.513 & 0.066 \\
			Estimates F & 2.722 & 1.290 & 0.647 & 0.079\\
			\noalign{\smallskip}\hline\noalign{\smallskip}
		\end{tabular}
	\end{center}
\end{table}

\noindent 
All the Bayesian estimators were computed using independent priors on the ($\beta, \sigma_u, \sigma_e$), 
with an improper flat prior on $\beta$, an Inverse Gamma with hyperparameters $(0.01, 0.01)$ on $\sigma^2_v$ and the Gelman's prior for $\sigma_u^2$.
In Table \ref{one1} results for the proposed estimators are reported. The values of estimates $A$ and $A^\ast$ are only affected by sample selection; the small differences between them can be explained in terms of sampling variability and the minimal effect of the Inverse Gamma prior over $\sigma_v$.

Estimates B are affected by missing matches: this only results in a sample size reduction due to non relevant bias in missing matches, at least in this simulated situation. On the other hand, the na\"ive estimates $C$ show the worst performance; this is mainly due to the introduction of false matches. As expected, this effect is well accounted for using the $D$ method. $C^\ast$ estimates are much better than their natural competitors $C$: this can be explained in terms of a better performance of the Bayesian Record Linkage in terms of point estimate of the matching matrix $C$. 
The proposed method $D$ produces a slight improvement when the magnitude of linkage errors is relatively low (the average in areas and replications is less than 15 \%). One can expect a more sensitive improvement with higher linkage error levels. 
The proposed adjustment is still subject to very restrictive assumptions, such as the identification of small areas with blocking variables in the linkage process, the exchangeability of linkage errors and, finally, the assumption of known linkage errors. When the vector $\bm{\lambda}$ (and $\bm{\gamma}$, if the exchangeability assumption is not postulated) need to be estimated, the trade-off of the adjustment between bias and variance should be assessed. 
In our simulation study, the $\lambda_d$'s were simply estimated as the relative frequency of corrects links. 
Among the three main assumptions described above, only the first one plays a role in the Bayesian approaches $E$ and $F$. We delay a general discussion of pros and cons of different methods to \S~\ref{five}. 
In terms of comparison between the two Bayesian strategies, one can see that, at least in our simulation set up, the general performance of the non-feedback effect strategy is definitely superior compared to that based on a feedback effect. 
We do not have a plain answer to explain this fact. Our conjecture is that the result may depend on the fact that the assumption of a linear relation between consumption and income, implicit in the unit level small area model, is not adequate for this data set.
In order to support our conjecture, we notice that, when the assumed model is ``correct'', the information contained in the variables involved in the small area model may contribute to flag the correct links. On the other hand, when the model is not correct, this advantage may turn itself into a bias, as in our example. 

The Bayesian approaches based on MCMC simulations also allow to provide an immediate estimate of the standard deviation of the estimators. Let us denote 
with $\hat{\sigma}(H, \beta)$ and $\hat{\sigma}(H, \alpha)$ the standard deviations of the posterior distribution of $\beta$ and $\alpha$ using method $H$. In our study we have obtained,
\begin{align*}
& \hat{\sigma}(E, \alpha)= 0.784; \quad \hat{\sigma}(E, \beta)=0.036 \\
& \hat{\sigma}(F, \alpha)= 0.596; \quad \hat{\sigma}(F, \beta)=0.025.
\end{align*}
 
Table \ref{author_tab:3} reports the absolute relative biases (ARB), the standard deviations and the MSE of all the competing estimators. ARB is defined as 
$$
\textnormal{ARB}= \frac 1D \sum_{d=1}^D \frac{\vert \hat{Y}_d - Y_d \vert} {Y_d }, 
$$
where $\hat{Y}_d$ is the predicted value of the consumption mean in area $d$, averaged over the 100 simulations 
and $Y_d$ is the true mean value.

\begin{table}
\begin{center}

\caption{Comparisons among estimators: ARB is the "\textit{Absolute relative bias}; SD is the observed standard deviation among different simulations; MSE is the mean square error. }
\label{author_tab:3}
\label{two}

\begin{tabular}{|p{2.3cm}|p{2.3cm}p{2.3cm}p{2.3cm}|}
\hline\noalign{\smallskip}
Estimates & ARB & SD & MSE \\
\noalign{\smallskip}\hline\noalign{\smallskip}

Estimates A  		& 0.033 & 0.463 & 0.517 \\
Estimates B  		& 0.033 & 0.540 & 0.595 \\
Estimates C  		& 0.043 & 0.508 & 0.705 \\
Estimates D  		& 0.035 & 0.523 & 0.605 \\
Estimates$A^{\ast}$ & 0.0286& 0.498 & 0.488 \\
Estimate$C^{\ast}$  &0.032  & 0.505 & 0.534 \\
Estimates E  		& 0.033 & 0.498 & 0.535 \\
Estimates F  		& 0.0289& 0.494 & 0.516 \\
Sample Mean 		& 0.0196& 1.908 & 4.297 \\
\noalign{\smallskip}\hline\noalign{\smallskip}
\end{tabular}

\end{center}
\end{table}

\noindent
On the other hand, one should also note from Table \ref{author_tab:3} 
that the estimation method $F$ performs better in terms of absolute relative efficiency: this may be due to a more accurate estimation of the random effects.

As a final comment on the simulation study, we notice that all methods behave sufficiently well; this happens because the key variables (apart from \texttt{Gender}) are really informative, with a large number of categories. 

We have also included the sample mean estimator among the competitors. 
From Table \ref{author_tab:3} one can notice how the sample mean outperforms all the proposed estimators in terms of bias; however, at the same time, it produces very large standard errors. 
The resulting mean square error of the sample mean is then very high; this confirms that, at the price of a possible increase in bias, composed and synthetic estimators may produce great benefits. Sample means show a relevant MSE mainly because in the dataset there are areas characterized by small sample sizes, as shown in Table \ref{NUMAREA}. 
A non standard case is represented by Area 17 which has a not so small sample size but it shows a large value of the MSE of the sample mean. 
This is likely due to the very high degree of variability of the consumption at population level in the above mentioned domain.
Another important point to stress is that, in our opinion, the increase in bias is mainly caused by an at least incomplete model specification and not by the linkage procedure.  In fact, in terms of bias, the sample mean outperforms also the benchmark estimators $A$ and $A^{\ast}$.  We argue that the model, being a very simple model between income and consumption, is not able to catch variability of $Y$.  

\section{Discussion}
\label{five}

The main objective of this paper was to compare different statistical methods to calibrate a unit level small area model in the presence of linked data.
Since the previous literature on this topic is relatively scarce, we have considered all the existing methods and compare them with some natural Bayesian versions of the same model. 

Given that a thoroughly comparison of the methods would imply an intensive simulation study, we confine ourselves in this paper to a practical comparison in a relatively typical situation as the one described in the previous section.
From a more general perspective, we stress the fact that the frequentist strategy $D$ can be rigorously implemented and a correction can be produced only when the exchangeability assumption holds. 
In practical situations, it is hard to meet an exchangeable structure of linkage errors; 
however, as in our simulation study, the "naive" application of estimator $D$ shows good a performance in a typical, probably not completely exchangeable, situation.
A drawback of the $D$ strategy may be found in the use of known values for $\bm{\lambda}$. A non reported sensitivity study shows that the final results are robust with respect to small variations of those value, although a more accurate sensitivity analysis should be considered. 

On the other hand, the Bayesian approaches $E$ and $F$ rely on minimal assumptions: the most important, which is common to all methods discussed here is that linkage errors may occur only within the same domain. Although this limitations can be avoided in theory, it is obvious that any linkage method must be based on some blocking mechanism in order to avoid computational intractability.

We should also say that we have confined ourselves to a comparative study in a situation where the key variable came from ``simple'' data sets, ready to be processed through standard record linkage procedure; we have not considered more complex situations because the main gist of the paper was the comparison between methods which can work reasonably well in standard applications of record linkage.

Another difference between frequentist and Bayesian approaches is the estimation of the vector $\bm{\lambda}$. While it represents one of the parameters to be routinely estimated in the Bayesian algorithm, his value is externally introduced when using method $D$: in these cases, $\bm{\lambda}$ can be either estimated through a training data set or using previous knowledge.

As far as a comparison within Bayesian methods is concerned, we believe that a feedback strategy should be preferred when a specific statistical model must be used and the model has been found adequate to fit the data.
In other situations, when the linkage process aims at producing a new data set which will be routinely used for many different purposes, then a non feedback strategy seems more appropriate.

Finally, our approach are essentially model-based, and their performance should always be considered in these terms.
When the model is not adequate, simple design based estimators may have a better performance, at least for moderate to large sample sizes.

\bibliographystyle{spbasic}

\section*{Acknowledgements}

\noindent 
The Authors warmly thank two anonymous referees and an Associate Editor whose comments greatly improved a previous version of this work. 
Andrea Tancredi's and Brunero Liseo's research was supported by the Italian Ministry of Education, PRIN 2015, grant number 2015EASZFS - PE1.

%


\end{document}